\documentclass[12pt]{article}
\usepackage{amsmath}
\usepackage{amssymb}
\usepackage[all]{xy}

\hoffset = - 1.cm
\voffset = - 1.2cm
\font\frak=eufm10 scaled\magstep1

 2

\newtheorem{theorem}{Theorem}
\newtheorem{prop}{Proposition}
\newtheorem{dfn}{Definition}

\newtheorem{lem}{Lemma}

\newtheorem{lem-theo}{Lemma-Theorem}

\newtheorem{dfn-prop}{Definition-Proposition}
  {\bigbreak\noindent{\bf Acknowledgements}\bigbreak}%
  {\bigbreak}
\newcommand{\beqn}{\begin{equation}}
\newcommand{\eeqn}{\end{equation}}
\newcommand{\beqnarray}{\begin{eqnarray}}
\newcommand{\eeqnarray}{\end{eqnarray}}
\def\pd#1#2{\frac{\partial #1}{\partial #2}}

\def\dfrac#1#2{{\displaystyle{\frac{#1}{#2}}}}
\def\<#1>{\langle#1\rangle}
\begin{document}

\title{Geometrical and Dynamical Aspects of Nonlinear Higher-Order
Riccati Systems \\}

\author{Jos\'e F. Cari\~nena$^{\dagger\,a)}$,
  Partha Guha$^{\ddagger\S\,b)}$, and
  Manuel F. Ra\~nada$^{\dagger\,c)}$     \\[8pt]
$^\dagger$ {\sl Departamento de F\'{\i}sica Te\'orica and IUMA } \\
      {\sl Universidad de Zaragoza, 50009 Zaragoza, Spain}   \\[3pt]
$^\ddagger$ {\sl  Centre Interfacultaire Bernoulli, EPFL}   \\
{\sl  SB CIB-GE, Station 7, CH-1015, Lausanne, SWITZERLAND }    \\[3pt]
$^\S$ {\sl S.N. Bose National Centre for Basic Sciences, JD Block}\\
{\sl  Sector-3, Salt Lake, Calcutta-700098, India} }

\date{}

\maketitle

\vspace{.01in}
\hspace{1.10in}

\begin{abstract}
We study a geometrical formulation of the
nonlinear second-order Riccati
equation (SORE) in terms of the projective vector field equation on $S^1$, which
 in turn is 
related to the stability algebra of Virasoro orbit.
Using Darboux integrability method we obtain the first-integral of SORE 
and the results are applied to the study of its Lagrangian and Hamiltonian
description. We  unveil the relation between the Darboux
polynomials and master symmetries associated to second-order Riccati.
Using these results we show the existence of a Lagrangian description for the
related system, and the Painlev\'e II equation is analysed.
\end{abstract}

\begin{quote}
%---------------

\bigskip


%%  34A26 Geometric methods in differential equations
%%  34A34  Nonlinear equations and systems, general
%%  35Q53  KdV-like equations (Korteweg-de Vries)

%---------------

%---------------
\end{quote}
{\vfill}

\footnoterule
{\noindent\small
$^{a)}${\it E-mail address:} {jfc@unizar.es}  \\
$^{b)}${\it E-mail address:} {partha@bose.res.in}  \\
$^{c)}${\it E-mail address:} {mfran@unizar.es}}
%---------------

\def\separador{\medskip{\hskip30pt\hrulefill\hskip30pt}\medskip}
\vfil\eject

%---------------
\tableofcontents
%---------------

%-----------------------------------------------

\section{Introduction }

The study of the  Riccati differential equations goes back to the early days of modern mathematical
analysis, since such equations represent one of the simplest
types of nonlinear ordinary differential equations and consequently 
Riccati equations play an important r\^ole in physics, mathematics and engineering sciences.
The usual first-order  Riccati equation appears as a reduction from a second-order linear ordinary differential equation when taking into account invariance under dilations according to Lie recipe (the inverse property is called the Cole--Hopf transformation). 
Such  correspondence between first-order  Riccati equations and second-order linear ODEs can be beautifully  manifested through their solutions. 
For example, if one solution of a linear second-order ODE is known, then it is known that any other solution 
can be obtained by means of a quadrature (i.e. a simple integration), and the same property holds true for the Riccati equation, but with two quadratures.
Moreover, first-order Riccati equation is a prototypical example of (systems of) differential equations  admitting a superposition rule, also called Lie systems \cite{CadL11, CarRam99}.  
This is why  some authors  \cite{Da62} have considered it as the first step in the study of nonlinear  (systems of) differential equations.

The second-order Riccati equation also becomes of paramount importance in recent years, mainly 
because of its connection with the integrable ODEs. It is known that the  second-order Riccati equations are 
related to a large class of integrable ordinary differential equations of anharmonic oscillators, viz. the Ermakov--Pinney, Normalised Ermakov--Pinney, Neumann type systems,
etc. \cite{BFL91,EEL07,Gu00Diff,Gu05IntJ,Gu2,Leach85}. 
It has been studied in \cite{CaRaS05} from a geometric perspective and it has been proved to admit two alternative Lagrangian formulations, both Lagrangians being of a non-natural class (neither potential nor kinetic term). 
A more geometric approach can be found in \cite{CaGR09} where  the theory of Darboux polynomials, the extended Prelle-Singer methods, (pre-)symplectic forms, or Jacobi multipliers, are used. 

Higher-order Riccati equations can be obtained by reduction from the Matrix Riccati equation \cite{HaWiA83,dORW}. 
Not only   first-order  Riccati equation but also higher-order Riccati differential equations can
be linearised via Cole-Hopf transformation (i.e.  they appear as a reduction from linear differential equations when taking into account  the dilation symmetry). 

The study of higher-order Riccati equations \cite{CaGR09} was also carried out 
from the perspective of  the theory of Darboux polynomials and the extended Prelle-Singer methods.
Higher-order Riccati equations
play the r\^ole of B\"acklund transformations for integrable
partial differential equations of order higher  than that of the KdV
equation. In fact Grundland and Levi \cite{GrunLev99} constructed the B\"acklund transformations
of several integrable systems from higher-order
Riccati equations.  
Such B\"acklund transformations are closely related to many important integrability
properties such as the inverse scattering method, the Painlev\'e
property, Lax pairs, and an infinite number of conservation laws \cite{Weiss}.

In the study of differential equations one finds cases that are
in some sense  solvable, or integrable, which enable one to study their 
dynamical behaviours using Lie theoretic methods. They often admit
different geometric and Hamiltonian formulations.
Integrable systems are a fundamental class of {\it explicitly solvable} 
dynamical systems of current interest in mathematics and physics.
One  notable example is the Milne--Pinney equation \cite{Ca08,CdL08,CLR08,Mi30,Pinney50}:
in 1950  Pinney presented  in  a one page paper \cite{Pinney50} the solution of the equation
$$
y'' + \Omega^2(x)y = \frac{1}{y^3}, 
$$
where the symbol $y''$ denotes the second derivative  with respect to the independent
variable $x$. He gave the solution in the form
$$
y(x) = (A\,\phi_{1}^{2} + 2B\,\phi_{1}\,\phi_{2} + C\,\phi_{2}^{2})^{1/2},
$$
where $y_1=\phi_1(x)$ and  $y_2=\phi_2(x)$ are any two linearly independent solutions of the associated linear equation $ y^{\prime \prime} + \Omega^2(x) y = 0$,
and  $A, B$ and $C$ are related according to $B^2 -AC = 1/W^2$ with
$W$ being the constant Wronskian of the two linearly independent solutions.
Ermakov introduced the above equation as an additional auxiliary equation  to a second order 
linear differential equation to define a system of two second order differential equations, and found  an invariant when multiplying by an integrating factor and obtained the invariant after integration
with respect to $x$.

The solutions of Painlev\'e's Second Equation (or Painlev\'e II)
\begin{equation}
y^{\prime \prime} = \alpha + xy + 2y^3\label{PainleveII}
\end{equation}
are meromorphic functions in the plane \cite{HL99,S00}.
It is known (see e.g. the paper by  Gromak \cite{Gromak} and \cite{GLS}) that every transcendental solution
$w$ has infinitely many poles with residue $+1$ and also with residue
$-1$, except when $\alpha = \pm 1/2$ and $w$ solves the Riccati equation (see \cite{S00})
$$\frac{dw}{dx}=\pm \frac x2\pm w^2.$$
In the paper we derive the Painlev\'e II, as well the second-order second
degree Painlev\'e II \cite{CosScou,JimboMiwa} using Riccati hierarchy, which is in turn connected to
the Virasoro stabilizer set or projective connection. The second-order second
degree Painlev\'e II is also known as Jimbo-Miwa equation \cite{JimboMiwa} and plays an important
role in random matrix theory \cite{ForWit01}.

%-----------------------------------------------

\subsection{Motivation and plan}

Our main goal is to study higher-order Riccati equations in various directions. 
In particular, we focus onto the geometrical aspects of the Riccati sequence. 
At first we explore its connection to the stabilizer set of the Virasoro orbit, also known as the projective vector field equation.
There are many papers devoted to Virasoro algebra and projective connection on $S^1$, but  very few papers discuss about their relations to nonlinear oscillator equations, Riccati chains and (second-order) Painlev\'e equations. 
This paper provides a relationship between the stabilizer set of the Virasoro orbit and second-order Riccati equations. 
Many papers (for example, \cite{Gu00Lett,Gu05JMAA,OvKh87}) have been written to elucidate the interaction between
the geodesic flows on the Bott-Virasoro group (see for example, \cite{KW08, Ki80, OT}) and integrable partial differential equations, especially the KdV type systems, but very few articles deal with finite-dimensional systems. 

 Some of the main questions  to be discussed in this paper can be summarized in the following points:
 
%---------------
\begin{itemize}

\item{} Relation of Riccati equation with the Virasoro algebra.

The Virasoro group plays a very important role in integrable systems.
It is known that the Virasoro group serves as the configuration space 
of the KdV and  the Camassa-Holm equations and these equations can be regarded 
as equations of the geodesic flows related to different right-invariant 
metrics on this group.  In other words, they have the same  symmetry group. 
Recall that the Virasoro group is a one-dimensional central extension of the group of smooth
transformations of the circle.  
In this work, we focus on the integrable dynamical systems connected to the 
stabilizer set of the Virasoro orbit. In fact the Milne--Pinney equation, which  describes the time-evolution
of an isotonic oscillator (also called pseudo-oscillator) -- i.e. an oscillator with
inverse quadratic potential--  is the first and foremost example of this class.  
It is known \cite{Gu00Lett,Gu05JMAA} that the entire (coupled)  KdV family is connected to the Euler-Poincar\'e formalism of the (extended) Bott-Virasoro group. 
This connection can be extended to the super Bott-Virasoro group \cite{GuOl06}.

The second-order Riccati equation has  also a nice geometric interpretation in terms of Virasoro orbit \cite{Gu05IntJ,Gu2}. 
Vector fields $f(x)\,{d}/{dx} \in {\mathfrak{X}}(S^1)$ associated to the stabilizer
set of Virasoro orbit are called  projective vector fields and
the corresponding differential equation satisfied by the function $f$ is called the projective vector
field equation  \cite{Gu02,Gu03}. It is known that the second-order Riccati
equations are associated to these equations. In fact, solutions of
a large class of $0+1$ dimensional integrable systems can be
expressed in terms of the global and local projective vector
fields.

\item{} Analysis of the infinitesimal symmetries of the  standard Riccati equation. 

This study is presented by making use of a geometric approach. It is proved that these symmetries are related to the projective vector field equation.

\item{} Study of the relation of the second-order Riccati equation with Painlev\'e equations.

There are  a certain number of properties relating the second-order Riccati equation with some other equations of the mathematical physics 
as the Airy equation or the Milne--Pinney equation \cite{Mi30,Pinney50}. Moreover, the second-order Riccati equation is also related to some of the Painlev\'e-Gambier equations  and this can be used to find the solutions of Painlev\'e's Second Equation (or  Painlev\'e II) for special values of parameters.

\end{itemize}
%---------------

The paper is organised as follows.  Section 2 is a short presentation of properties of first- and higher-order Riccati equations and their relations with linear equations and nonlinear superposition rules.
In Section 3 we give a concise introduction to Virasoro orbit, stabilizer set and projective connections and in Section 4 we study the infinitesimal symmetries of standard Riccati equation and their relations with  the projective vector field equation.
We elucidate in Section 5 the connection between the ordinary Painlev\'e II and second-order second degree Painlev\'e II 
equations with the Virasoro orbit in Section 5. 
%-----------------------------------------------

\section{First-order and higher-order Riccati equations}

Our main goal is to explore the higher-order Riccati equations, but before giving  the formal description  of  higher-order Riccati equations and Riccati chain, a little introduction may be helpful. 
So we start our journey by giving a short introduction to  first-order Riccati equation.

The usual Riccati equation
\begin{equation}\label{Ric1}
 u^{\prime} = f(x) + g(x)u + h(x)u^2,
\end{equation}
is a  first-order nonlinear differential equation with a quadratic non-linearity. 
The solutions of the Riccati equation are free from movable branch points and can have only movable poles \cite{AblFok,Hille,Ince44}.
This is the simplest nonlinear differential equation admitting a superposition
rule for expressing the general solution in terms of particular solutions.
 The linear fractional transformation (or change of variables)
 \beqn \bar u = \frac{a(x)u + b(x)}{c(x)u+ d(x)}\,, \qquad ad-bc=1\,, \eeqn
transforms each Riccati equation into another one \cite{CMN98}. In order to
solve a Riccati equation by quadratures, it is enough to know one particular
solution which allows us to reduce the problem to carry out two
quadratures, while when two particular solutions are known the
problem can be reduced to a new differential equation solveable by   just one quadrature, and
finally if three particular solutions $u_1(x), u_2(x), u_3(x),$ are known,
we can construct all other solutions $u$ without use of any
further quadrature \cite{CarRam99}. This is carried out by using the
property that the cross-ratio of four solutions is a constant
(i.e. for any other solution $u(x)$, there is a  real number  $k$ such
that: 
\beqn 
  \frac{u(x) - u_1(x)}{u(x) - u_2(x)} = k \,\frac{u_3 (x)- u_1(x)}{u_3(x) - u_2(x)},
\eeqn 
where $k$ is an arbitrary constant characterising each
particular solution. For instance $u_1(x)$ corresponds to $k=0$ and $u_2(x)$ 
is obtained in the limit of $k\to \infty$).

It was shown by Lie and Scheffers \cite{Lie1893} that the Riccati
equation is essentially the only first-order  ordinary nonlinear differential equation of
which possesses a nonlinear superposition rule which  comes from the preceding relation:
$$
u=\Phi(u_1,u_2,u_3;k)=\frac {u_1(u_3-u_2)+ku_2(u_1-u_3)}{u_3-u_2+k(u_1-u_3)},
$$ 
(i.e. if $u_1(x), u_2(x)$ and $u_3(x)$ are solutions of (\ref{Ric1}), then for any real value  $k$,
$$
  u(x) = \frac{u_1(x) (u_3 (x) - u_2(x) ) + ku_2(x) (u_1(x)  - u_3(x) )}{u_3(x)  - u_2(x)  + k(u_1(x)  - u_3(x) )},
$$
is another solution and each solution is of this form for an appropriate choice of $k$).

Therefore the general solution $ u(x)$ is non-linearly expressed
in terms of three generic particular solutions, $u_1(x)$, $u_2(x)$ and  $u_3(x)$, and  a
constant parameter $k$.

Remark also that a first-order Riccati  differential equation (\ref{Ric1}) with $h(x)$ of constant sign in an open set, transforms under the relation 
\begin{equation}
u=-\frac 1h \frac{y'}{y}\label{hCH}
\end{equation}
into a second order differential equation 
\begin{equation}
y''+ a_1(x)y'+a_0(x)y=0\label{lsode}
\end{equation}
with 
\begin{equation}
a_1=-g-\frac{h'}h,\qquad a_0= f\,h. \label{coefl2vR}
\end{equation}
This relation (\ref{hCH}) is not a change of coordinates or transformation but for any real number $\lambda$, $y$ and $\lambda \, y$ have associated the same function $u$.
Conversely, given the  linear second-order differential equation (\ref{lsode}), for any nowhere vanishing function $h$, (\ref{hCH}) reduces  (\ref{lsode}) to the Riccati equation (\ref{Ric1}) with 
coefficients determined by(\ref{coefl2vR}).

In a similar way,  consider the following nonlinear second-order equation in an open set $I$ \cite{Da62}:
\begin{equation}
 u'' + \bigl[\,\beta_0(x) + \beta_1(x) u\,\bigr] u' +
     \alpha_0(x) + \alpha_1(x) u + \alpha_2(x) u^2 + \alpha_3(x) u^3 = 0 \,, \quad x\in I\subset \mathbb{R},
\label{Riccati2}
\end{equation}
where we suppose that $\alpha_3(x)>0, \forall x\in I $ and the two functions $\beta_0$, $\beta_1$,
are not independent but satisfy
$$
\beta_0 = \frac{\alpha_2}{\sqrt{\alpha_3}} - \frac{\alpha_3'}{2 \alpha_3} \,,\quad
\beta_1 = 3 \sqrt{\alpha_3} \,.
$$
In particular, the case $\beta_0=\alpha_1= \alpha_2=0$, $\alpha_3=1$,  $\beta_1=3$,  was studied in Davis and Ince books \cite{Da62,Ince44}
and will be more carefully considered in next sections, while that of   $\beta_0=\alpha_0=\alpha_1= \alpha_2=0$,  $\beta_1=1$,  was
studied by Leach and coworkers \cite{Leach85,LFB88} who proved that only the case $\alpha_3(x)=1/9$ is linearizable, possesses eight symmetries and is completely
integrable. 

An important property is that, as indicated in \cite{CaRaS05}, the nonlinear equation (\ref{Riccati2}) can be transformed into a third-order linear equation
\begin{equation}
 y''' + a_1(x) y'' + a_2(x) y' + a_3(x) y  = 0\label{ltode}
\end{equation}
by the substitution 
\begin{equation}
  u(x) = \frac{1}{\sqrt{\alpha_3(x)}}\,\frac{y'(x)}{y(x)} \,. \label{aCH}
\end{equation}

The equation (\ref{Riccati2}) is a second-order generalization
of the usual first-order  Riccati equation and because of this is usually known
as the Riccati equation of second-order  \cite{CaRaS05}:
it is   a nonlinear equation whose solution of can be 
expressed in terms of solutions of a  third order linear differential equation. 
Conversely, if we start with a linear third-order equation then we can get the associate nonlinear second-order equation by introducing the appropriate reduction of order.
 
One can go further and  define  $j$-order Riccati equations as those  appearing  as a reduction of a $(j+1)$-order linear differential equation  \cite{CaGR09}.
 More specifically, given a differential equation $y^{(n)} =d^ny/dx^n=0$, its invariance under dilations suggests, according to Lie recipe, to look for
  a new variable $z$ such that the dilation vector field $y\, \partial /\partial y$ becomes
$(1/k) \partial /\partial z$, the factor $k$ being purely
conventional (we often put $k=1$). 
Then $y=e^{kz}$, up to an irrelevant factor, and with this change of variable, $y^{(n)}=0$, for $n>1$,  becomes $R^{n-1}(u)=0$ with $u=\dot z$  and $R$ the differential operator $R=(D+k\, u)$.
In fact, when $D=d/dx$, we have that  $e^{-kz}De^{kz}=D+k\, u$, and then we can prove by complete induction that $y^{(n)}=k\, e^{kz}R^{n-1}(u)$ (see \cite{CaGR09}):
the property is true for $n=1$, because $y'=k\,\dot z\,
e^{kz}= k\,e^{kz}\,u=k \, e^{kz}\, R^{0}(u)$, and if the
property is true till  $n=j$, it also holds for $n=j+1$, because
$$
  D^{j+1}y=D(y^{(j)})=D(k\,e^{kz}R^{j-1}(u))=k\,e^{kz}(D+k\,
u)R^{j-1}(u)=k\,e^{kz}\,R^{j}(u).
$$
Therefore, Lie recipe applied to the invariant under dilations equation $y^{(n)}=0$ transforms such an equation  into
$R^{(n-1)}(u)=0$. 
The first terms of such Riccati sequence are 
\begin{equation}
R^0(u)=u,\qquad R^1(u)=u'+k\,u^2, \qquad R^2(u)=u''+3k\,u\, u'+ k^2u^3,\qquad \ldots \label{firstRiceqs}
\end{equation}
\begin{equation}
R^3(u)=u''' + 4u\,u'' + 3u^{\prime 2} + 6u^2\,u' + u^4 ,\ R^4(u)=u^{(iv} + 5u\,u''' + 10u'u'' + 15uu^{\prime 2} + 10u^2u'' + 10u^3u' + u^5,  \ldots \label{moreRiceqs}
\end{equation}
and for the  general linear differential equation of order $n$,
\begin{equation}
 a_0(x)y+\sum_{j=1}^na_ j(x)\, y^{(j)}=0, \quad a_n(x)=1,  \label{genhoRiceq}
\end{equation}
reduces to
$$ 
 a_0(x)+ \sum_{j=1}^na_ j(x)\,R^{j-1}(u)=0. 
$$

%-----------------------------------------------

\section{Virasoro algebra, projective vector field and second order Riccati equaltion}

Let $\Omega^1=T^*S^1$ be the cotangent bundle of a circle $S^1$. Since $S^1$ is 
 diffeomorphic to the 1-dimensional Lie group $U(1)$, there exists a canonical coordinate $x$, 
 defined up to a factor, which we can fix such that the domain of the chart is $(0,2\pi)$.
This $\Omega^1$ is a trivial real line bundle on $S^{1}$. Similarly, let  $\Omega^m$ denote  the $m$-fold tensor
product of $\Omega^1$. The local coordinate expression of a  section of $\Omega^m$ is given by $s(x) = {g}(x)\, dx^m $, 
where $dx^m $ is a shorthand notation for the tensorial product of $m$ times $dx$.  
The set  $\Gamma (\Omega^m)$ of such sections is a free $C^\infty(S^1)$-module, a natural basis being given by $dx^m$, 
and in this sense such sections are described by functions in $S^1$. 
A vector field in $S^1$ is of the form 
\begin{equation}
{X_f} = f(x)\frac{d}{dx} \in {\mathfrak{X}}(S^1),\label{vfXf}
\end{equation}
where $f$ is a function on $S^1$ (i.e. a $2\pi$--periodic function on the real line).
Taking into account that 
$$
{\mathcal{L}}_{X_f} dx^{m}=m\,f^{\prime}\,dx^{m}
$$
we see that the infinitesimal action of  $X_f$ on  a section $s=g(x)\, dx^m$ of $\Omega^m$ is given by (see e.g. \cite{Gu02})
\beqn 
{\mathcal{L}}_{X_f} s =
(fg^{\prime} + mf^{\prime}g )dx^{m}, \label{action}
\eeqn 
where ${\mathcal{L}}_{X_f}$ is the Lie derivative with respect to the vector field $ {X_f}$ given by (\ref{vfXf}). 

Sometimes it is convenient to use the above mentioned identification of a section with its component, a function, 
and then the Lie derivative when acting on sections of  $\Omega^p$ expressed in terms of functions,
denoted ${\mathcal{L}}_{X_f}^{(p)}$,  is defined as
\beqn
{\mathcal{L}}_{X_f}^{(p)} = f(x)\frac{d}{dx} + pf^{\prime}(x).\label{LXpsect}
\eeqn
This really means that if  $s=\psi(x)\, dx^p$, then
$$
{\mathcal{L}}_{X_f}s=\left(f(x)\frac{d\psi}{dx} + pf^{\prime}(x)\,\psi(x)\right)\,dx^p 
 ={\mathcal{L}}_{X_f}^{(p)}\psi\ dx^p.
$$

There is a natural symmetric bilinear map $\Gamma (\Omega^{m_1})\times \Gamma (\Omega^{m_2})\to \Gamma (\Omega^{m_1+m_2})$
 given by the usual commutative product of tensor densities
$$
f(x)(dx)^{m_1} \otimes g(x) (dx)^{m_2} \mapsto f(x)g(x) (dx)^{m_1+m_2}.
$$
In particular  for $m=0$,  $\Gamma (\Omega^{0})=C^\infty(S^1)$. We can at least formally extend the values of the index $m$  to the set of integer numbers, or even to rational numbers,  and then the product by elements of $ \Gamma (\Omega^{0})$ is the external composition law  of the module structure. Note that the elements of  $\Gamma (\Omega^{-1})$ when multiplied by those of $\Gamma (\Omega^{m})$ play as an inner contraction with a vector field and in this sense $\Gamma (\Omega^{-1})$ is to be identified with  ${\mathfrak{X}}(S^1)$ (i.e. 
we can  denote the tangent bundle as
$\Omega^{-1}$ ($\equiv TS^1$)).

 Note also that when  $X_g = g(x){d}/{dx} \in \Omega^{-1}$, then putting $p=-1$ in  (\ref{LXpsect}) we get the usual expression:
\beqn
{\mathcal{L}}_{X_f}{X_g} = \left[ f(x)\frac{d}{dx}, g(x)\frac{d}{dx}\right]= (f\, g^{\prime} -g\, f^{\prime} ) \frac{d}{dx} .
\eeqn

We will denote $\Omega^{\pm 1/2}$  the `square root' of the tangent
and cotangent bundle of $S^1$ respectively. The respective sections of such bundles will be of the form
$s= \psi (x)dx^{\pm\frac{1}{2}} \in \Gamma (\Omega^{\pm\frac{1}{2}})$.

Another remarkable property is that the $C^\infty(S^1)$-module $\Gamma (\Omega^m)$ of  sections of the $m$-fold 
tensor product bundle $\Omega^m$ has a natural structure of a Poisson algebra with the 
commutative product given by the above usual product of tensor densities
and the Poisson algebra commutator given by the  Rankin-Cohen bracket
$$
\big\{f(x)(dx)^m,g(x)(dx)^n \big\} = \big(mf(x)g^{\prime}(x) - n f^{\prime}(x)g(x) \big)(dx)^{n+m+1},
$$
which is also known as first transvectant.

The group  ${\rm Diff\,}(S^1)$ acts on $\Gamma(\Omega^m)$, as given by 
$$\Phi^{\ast}(f(x)(dx)^m ) = f(\Phi(x))(\Phi^{\prime}(x))^{m}(dx)^m  $$
for $\Phi \in {\rm Diff\,}(S^1)$. The Lie algebra of such group is identified with 
the Lie algebra ${\mathfrak{X}}(S^1)$ and acts on this space $\Gamma(\Omega^m)$ by the Lie derivative given in (\ref{action}).

The theory of projective connections on the circle has much to do with the theory of Riccati equations we are dealing with. 
We will use the following definition of a projective connection  \cite{Hit91}:
%-------------------
\begin{dfn}
A projective connection on the circle is a linear second-order differential operator
\begin{equation}
  \Delta : \Gamma (\Omega^{-\frac{1}{2}}) \longrightarrow
\Gamma (\Omega^{\frac{3}{2}})
\end{equation}
such that: 
\begin{enumerate}
\item the principal symbol of $\Delta$ is the identity and
\item ${\displaystyle\int_{S^1}} (\Delta s_1)s_2 ~=~ {\displaystyle\int_{S^1}} s_1 (\Delta s_2)$, \
 for  all pairs of sections $s_i \in \Gamma (\Omega^{-\frac{1}{2}}).$
\end{enumerate}
\end{dfn}

Let us take $ s = \psi (x)dx^{-\frac{1}{2}} \in \Gamma (\Omega^{-\frac{1}{2}})$,
then a  linear second-order differential operator is such that  $\Delta s \in \Gamma (\Omega^{3/2})$ is locally described by \cite{Hit91}
\beqn 
 \Delta s ~=~ (a\, \psi^{\prime \prime} + b\, \psi^{\prime} + c\,\psi )\,dx^{\frac{3}{2}}, 
\eeqn
where $a,b$ and $c$ are real numbers.

The first condition in the definition of the projective connection,  implies that for $\Delta$ to be a projective connection it must be $a = 1$, and, on the other side, the second condition
implies that $b = 0$. Hence each projective connection can be identified with a Hill operator 
\beqn  
  \Delta_v ~=~ \frac{d^2}{dx^2} + v(x),
\eeqn
where $v$ is an arbitrary function in $S^1$ (i.e  a periodic function in $\mathbb{R}$).

%-------------------
\begin{dfn} Given a projective connection
\begin{equation}
\Delta_v = \frac{d^2}{dx^2} + v(x),\label{pcv}
\end{equation}
a vector field $X_f \in {\mathfrak{X}}(S^1)$ is called projective
vector field with respect to the 
projective connection $\Delta_v $ when it leaves  invariant the  projective connection (i.e.
\beqn  {\mathcal{L}}_{X_f}^{(3/2)} {\Delta}_vs ~=~ \Delta_v ({\mathcal{L}}_{X_f}^{(- 1/2)} s),\label{invpc}
\eeqn for all $ s \in \Gamma
(\Omega^{-\frac{1}{2}})$, where ${\mathcal{L}}_{X_f}^{(p)}$ is the expression for the Lie derivative
with respect to $X_f$ given in (\ref{LXpsect})).
\end{dfn}

The characterization of projective vector fields with respect to a given
projective connection is given in the following theorem (see \cite{Gu02}):
 
%-------------------
\begin{theorem}
$X_f\in \Gamma (\Omega^{-1})$ is  a projective vector field with respect to the 
projective connection (\ref{pcv}) if and only if $f$ 
 satisfies  (see \cite{Gu02}) 
 \beqn f^{\prime \prime \prime} +
4v\,f^{\prime} + 2v^{\prime}\,f ~=~ 0. \label{pvfc} 
\eeqn
-- i.e. $f$ is a solution of  the third-order linear differential equation called projective vector field equation:
\beqn  
 y^{\prime \prime \prime} + 4v\,y^{\prime} + 2v^{\prime}\,y ~=~ 0. \label{pvfe} 
\eeqn
\end{theorem}
{\it Proof}: In fact, if $ s = \psi (x)dx^{-\frac{1}{2}} \in \Gamma (\Omega^{-\frac{1}{2}})$, then 
$$\Delta_vs=(\psi''+v\,\psi) \,dx^{3/2},
$$
and therefore,
$${\mathcal{L}}_{X_f}^{(3/2)} {\Delta}_vs =\left[f(\psi'''+v'\,\psi+v\,\psi')+\frac 32 f'(\psi''+v\,\psi)\right]\,dx^{3/2}
$$
On the other side,
$$ 
{\mathcal{L}}_{X_f}^{(-1/2)} s=\left(f\,\psi'-\frac 12 f'\,\psi\right) \, dx^{-\frac{1}{2}} ,
$$
and thus,
$$\Delta_v({\mathcal{L}}_{X_f}^{(-1/2)} s)=\left[\left(\frac{d^2}{dx^2}+v\right)(f\,\psi'-\frac 12 f'\,\psi)\right]\,dx^{3/2}
$$
$$=\left(
f\,\psi'''+\frac 32 f'\,\psi''+v\,f\, \psi'-\frac 12 (v\,f'+f''')\psi\right)\,dx^{3/2}
$$
Consequently,
$$
{\mathcal{L}}_{X_f}^{(3/2)} {\Delta}_vs -\Delta_v({\mathcal{L}}_{X_f}^{(-1/2)} s)=\left(v'\,f+\frac 32v\,f'+\frac 12 f'''+\frac 12 v\,f' \right)\psi\, dx^{3/2},
$$
from where  we see that the invariance condition (\ref{invpc}) implies (\ref{pvfe}).

Each solution of the  projective vector field equation provides a local projective vector field, while a global solution of the equation, with the mentioned periodicity condition defines a global 
projective vector field. 

   It will be  proved in next subsection that these projective vector fields can alternatively be seen to be  the elements generating the stability subalgebra  
   of the point $(-1,v\,dx^{2})$ of the   Virasoro orbit of the corresponding $v$ (see next).  
   
   Let us remark that sometimes $v$ in (\ref{pcv}) is replaced by $k\, v$, where $k$ is a real number,and then equation (\ref{pvfc}) becomes:
\begin{equation}
   f^{\prime \prime \prime} +4k\,v\,f^{\prime} + 2k\,v^{\prime}\,f ~=~ 0.
    \label{pvfck} 
\end{equation}
 -- i.e. $f$ is a solution of the third-order linear differential equation 
\begin{equation}
   y^{\prime \prime \prime} +4k\,v\,y^{\prime} + 2k\,v^{\prime}\,y ~=~ 0.
    \label{pvfek} 
\end{equation}

%-----------------------------------------------
 
\subsection{Virasoro algebra and projective vector field equation}

Let ${\rm Diff}_+(S^1)$ be the group of orientation preserving diffeomorphisms of the circle $S^1$. 
We represent an element of ${\rm Diff}_+(S^1)$ as a diffeomorphism $\Phi(e^{i\,x})=e^{if(x)}$ where 
$ f :\mathbb{R} \longrightarrow \mathbb{R}$ is a function such that
(a)\, $f \in C^{\infty}(\mathbb{R})$, (b)\, $f(x+2\pi) = f(x) + 2\pi$, (c)\, $f^{\prime}(x) > 0$.
Therefore the  tangent space $T_{{\rm id}}{\rm Diff}_+(S^1)$ is the set of elements  $f(x)\,{d}/{dx} \in {\mathfrak{X}}(S^1)$ with $f$ a periodic differentiable function such that $f^{\prime}(x) > 0$. Not the however that under a change of coordinates $f$ does not change as a function but as a vector coordinate.

The group ${\rm Diff}_+(S^1)$  is endowed with a smooth manifold structure based 
on the Fr\'echet space $C^{\infty}(S^1)$. Its Lie algebra $\mathfrak{g}$ is the real linear
space  ${\mathfrak{X}}(S^1)$.
It is known that the group ${\rm Diff}_+(S^1)$ has non-trivial one-dimensional
central extensions by $U(1)$, the Bott-Virasoro group ${\widehat {\rm Diff}_+(S^1)}$. 
Recall that a  central extension is given by an exact sequence of groups 
$$
\xymatrix{ 1\ar[r]&U(1)\ar[r]^j&{\widehat {\rm Diff}_+(S^1)}\ar[r]^\pi&{\rm Diff}_+(S^1)\ar[r]&1,}
$$
where $j(U(1))$ lies in the centre of ${\widehat {\rm Diff}_+ (S^1)}$. Choosing a normalized  (local) section $\xi$ for $\pi$
we see that $\xi(\sigma_1 \circ \sigma_2)$ differs from the product of  $\xi(\sigma_1)$ and  $\xi(\sigma_2)$ in an element  $c(\sigma_1 , \sigma_2)$ of $j(U(1))$.  
The associativity is equivalent to the cocycle condition
$$
c(\sigma_1 \circ \sigma_2,\sigma_3)+c(\sigma_1 , \sigma_2) = 
 c(\sigma_1, \sigma_2 \circ \sigma_3)+c(\sigma_2,\sigma_3).
$$
Changing the section amounts to modify the cocycle $c$ by a coboundary $\zeta$ (i.e. a cocycle for which  there exists a map  $\tau:{\rm Diff}_+ (S^1)\to U(1)  $
 such that $\zeta(\sigma_1 , \sigma_2)=\tau(\sigma_1 \circ \sigma_2)\circ (\tau(\sigma_1 ))^{-1}\circ (\tau(\sigma_2 ))^{-1}$).

 More explicitly, the Bott-Virasoro group  ${\widehat {\rm Diff}_+ (S^1)}$ is the central extension
 of ${\rm Diff}_+(S^1)$ by $U(1)$ determined  by the  {\it Bott cocycle} \cite{Gu05JMAA}
 $$  
  c : \hbox{   } {\rm Diff}_+(S^1) \times {\rm Diff}_+(S^1) \longrightarrow \mathbb{R} 
 $$
 given by
 $$  
 c(\sigma_1 , \sigma_2)  = \frac 12 \int_{S^1}\hbox{  }\log(\sigma_1 \circ \sigma_2)^{\prime}\ 
 d\log |\sigma_{2}^{\prime}|,
$$
 for $\sigma_i \in {\rm Diff}_+(S^1)$. This cocycle satisfies the normalization condition
 $c(\sigma_1 , \sigma_{1}^{-1}) = 0 $. The cocycle can also be written as 
$$  
 c(\sigma_1 , \sigma_2)  = \frac 12 \int_{S^1}\ \log(\sigma_1^{\prime} \circ \sigma_2)\ 
 d\log |\sigma_{2}^{\prime}|,
$$
because using the chain rule we have $(\sigma_1 \circ \sigma_2)'=(\sigma_1^{\prime} \circ \sigma_2)\,\sigma_2^\prime$ and then, 
$$ 
 \frac 12 \int_{S^1}\ \log(\sigma_1 \circ \sigma_2)^{\prime}\ 
 d\log |\sigma_{2}^{\prime}|=\frac 12 \int_{S^1}\ \log(\sigma_1^{\prime} \circ \sigma_2)\ 
 d\log |\sigma_{2}^{\prime}|+\frac 12 \int_{S^1}\ \log(\sigma_2')\ d\log |\sigma_{2}^{\prime}|
$$
and integration by parts shows that the last term vanishes as a consequence of periodicity.
 
 Note that 
$$c(\sigma_1 \circ \sigma_2,\sigma_3)=\frac 12 \int_{S^1}\log(\sigma_1 \circ \sigma_2\circ \sigma_3)^{\prime}\ 
 d\log |\sigma_{3}^{\prime}|=\frac 12 \int_{S^1}\log(\sigma_1^{\prime} \circ \sigma_2\circ \sigma_3)\ 
 d\log |\sigma_{3}^{\prime}|+c(\sigma_2,\sigma_3)
 $$
 and similarly,
 $$
 c(\sigma_1 , \sigma_2\circ\sigma_3)=\frac 12 \int_{S^1}\log(\sigma_1 \circ \sigma_2\circ \sigma_3)^{\prime}\ 
 d\log |\sigma_2\circ\sigma_{3}^{\prime}|=\frac 12 \int_{S^1}\log(\sigma_1^{\prime} \circ \sigma_2\circ \sigma_3)\ 
 d\log |\sigma_{3}^{\prime}|+c(\sigma_1,\sigma_2)
 $$
 from where the cocycle condition follows. If $(t,\sigma)$ denotes the product $j(t)\xi(\sigma)$ the composition law in ${\widehat {\rm Diff}(S^1)}$ is 
 $$ 
 (t_1,\sigma_1)\cdot (t_2,\sigma_2)=(t_1+t_2+ c(\sigma_1 , \sigma_2), \sigma_1 \circ\sigma_2).
 $$
 
 The corresponding  non-trivial central extension of the Lie algebra ${\mathfrak{X}}(S^1)$ 
by the trivial Lie algebra   $\mathbb{R}$ is called the Virasoro algebra and denoted ${\mathfrak{vir}}$
-- i.e. we have a central extension 
 $$ 
 \xymatrix{ 0 \ar[r]& {\mathbb{R}}\ar[r]& {\mathfrak{vir}}\ar[r]&
 {\mathfrak{X}}(S^1) \ar[r]& 0.} 
 $$  
More specifically,  the elements of ${\mathfrak{vir}}$ can be identified with pairs 
(\,real number,$2\pi$-periodic function). In other words, the $2\pi$-periodic function is the component of an element of  the set ${\mathfrak{X}}(S^1)$, which is known to be  endowed with a Lie algebra structure.
The corresponding cocycle $\zeta:{\mathfrak{X}}(S^1)\times {\mathfrak{X}}(S^1)\to \mathbb{R}$ is given by 
$$ 
 \zeta(X_f,X_g)=\int_{S^1}f^{\prime}g^{\prime \prime}\; dx,
$$
where $X_f=f(x)\,\partial/\partial x$ and  $X_g=g(x)\,\partial/\partial x$. 
Obviously, $\zeta(X_f,X_g)=-\zeta(X_g,X_f)$ as a simple integration by parts shows, because of the periodicity of functions $f$ and $g$.

In fact, the corresponding cocycle will be given by
$$ 
 \zeta(X_f,X_g)=\frac{d^2}{dt\, ds}\left(c(\phi_t,\varphi_s)\right)_{|s=t=0}-
\frac{d^2}{dt\, ds}\left(c(\varphi_s,\phi_t)\right)_{|s=t=0},
$$
where $\phi_t$ and $\varphi_s$ are the flows of $X_f$ and $X_g$, respectively.

Taking into account that 
$$
\frac{d}{dt}\left(c(\phi_t,\varphi_s)\right)_{|t=0}=\frac 12 \int_{S^1}\ (\log '(\phi_0\circ \psi_s)^\prime)(f\circ\varphi_s)' d\log\varphi'_s=\frac 12 \int_{S^1}\
(f'\circ\varphi_s) d\log\varphi'_s,
$$
and
$$\frac d{ds}\left(\frac 12 \int_{S^1}\
(f'\circ\varphi_s) d\log\varphi'_s\right)_{s=0}=\frac 12 \int_{S^1}\ f'\, dg',
$$
and similarly for the other term, we find that 
$$ 
 \zeta(X_f,X_g)=\frac 12\int_{S^1} (f'\, dg'-g'\, df')=\int_{S^1} \,f^{\prime}g^{\prime \prime}\; dx.
$$

Therefore the  commutator in ${\mathfrak{vir}}$ takes the form
$$ 
\displaystyle \Big[ (a,X_f), (b,X_g) \Big] =(\zeta(X_f,X_g),[X_f,X_g])
= \left( \int_{S^1}\ f^{\prime}g^{\prime \prime}\; dx, (fg^{\prime} - gf^{\prime})\frac{d}{dx} \right). 
$$

The dual linear space ${\mathfrak{vir}}^{\ast}$ can be identified to the
set $\{ (\mu, v\,dx^2)\mid  \mu \in \mathbb{R}, v\in C^{\infty}(S^1)\}$.  
In fact, an element  $(\mu, v\,dx^{2})$ maps linearly  ${\mathfrak{vir}}$ into the set of  the real numbers as follows: 
$$  
 \< (\mu, v\ dx^2), (a,X_f) > = a\, \mu + \int_{S^1}f(x)\,v(x)\ dx, 
$$ 
and conversely, each linear map from   ${\mathfrak{vir}}$ into $\mathbb{R}$  can be represented as such a pair $(\mu, v\ dx^2)$.

The remarkable point is that if ${\rm ad\,}_{(a,X_f) }$ is the image under the  adjoint representation of 
the element $(a,X_f) $ of the Virasoro algebra  ${\mathfrak{vir}}$ and ${\rm ad\,}_{(a,X_f) }^{\ast}$ is the adjoint element, then
$$
 {\rm ad\,}_{(a,X_f)}^{\ast} (1, v\ dx^2)  = \left( 0, \frac{1}{2}f^{\prime \prime \prime}
 + 2v\,f^{\prime} + v^{\prime}\,f\right). 
$$

 In fact, it follows  from the definition
 $$\begin{array}{rcl}
  \< {\rm ad\,}_{(a,X_f) }^{\ast} (\mu, v\ dx^2), (b,X_g) >
&=& \< (\mu, v\ dx^2), {\rm ad\,}_{(a, X_f)}(b,X_g) > \\
&=& \left\langle (\mu, v\ dx^2), \left({\displaystyle\int}_{S^1} 
f^{\prime}g^{\prime \prime} dx, [X_f,X_g]\right) \right\rangle\\ & &\\
&=&{\displaystyle{\mu} \!\!                                                                                                                                                                                                                                                                                                                                                                                                                                       \int_{S^1}f^{\prime}g^{\prime \prime} \!\! + \!\! \int_{S^1}}v(fg^{\prime} - f^{\prime}g)dx= \!\! {\displaystyle\int_{S^1} \!\! \left( \mu f'''-2v\,f'-v'f\right)}g(x) dx,
\end{array}
$$
and then 
$$
{\rm ad\,}_{(a,X_f) }^{\ast} (\mu, v\ dx^2)=\left(0, \mu\, f'''-2v\,f'-v'\,f\right).
$$

Hence we see that  the stability Lie algebra of the point $(-1/2, v\ dx^2)$ 
relative to the action defined by  ${\rm ad\,}^{\ast}$ 
on its dual is given by an  $X_f \in {\mathfrak{X}}(S^1)$ such that $f$ is a 
solution of the third order differential equation (\ref{pvfe}).

%-----------------------------------------------

\subsection{Projective vector field equation and its structure}

Equation (\ref{pvfe}) is a linear equation and it is  therefore invariant under 
the dilation vector field. Lie recipe amounts to introduce a new dependent variable $z$
instead of $y$ in such a way that the dilation vector field has the form
$D=\partial{}/\partial z$ (i.e. $y=C e^z$, for any constant $C$),  and then 
\begin{equation}
y'=Cz'\, e^z\,,\quad y''=C(z''+z^{\prime 2})e^z \,,\quad y'''=C
(z'''+3z' z''+z^{\prime 3})e^z\,.\label{dilchange}
\end{equation}
Note that $u=z'=y'/y$, no matter of the value of $C$.
The important point is that the transformed equation does not depend on $z$ but on its derivatives and the
order of the equation is reduced by one just introducing the variable $u=z'$ and so we obtain
\begin{equation}
 u''+3u\,u'+u^3+4vu+2v'=0\,. \label{projReqk1}
\end{equation}

The purpose of this section is to set up a geometrical 
formulation of such nonlinear second-order ODEs in terms of the
projective vector field equation. 

%---------------------

\begin{prop}
Let $\psi_1$ and $\psi_2$ be two linearly independent solutions of the second order 
differential equation 
\beqn
\psi^{\prime \prime} +k\, v\psi =0 \,. 
 \label{Eqkv2prop1}
\eeqn
 Then,  the three-dimensional linear space of solutions of the third order equation 
\beqn
 y^{\prime \prime \prime}+ 4k\,v\,y^{\prime} + 2k\,v^{\prime}\,y  = 0
 \label{Eqkv3prop1}
\eeqn
is spanned by the functions $\psi_{1}^{2}$, $\psi_{2}^{2}$, and $\psi_{1}\psi_{2}$.
\end{prop}

{\it Proof}:
If $\psi_1$ and $\psi_2$ are two linearly independent  solutions of the equation (\ref{Eqkv2prop1})
then taking derivatives we obtain
$$ 
 \psi_1'''+kv\,\psi_1'+kv'\,\psi_1 = 0 \,,{\qquad} 
 \psi_2'''+kv\,\psi_2'+kv'\,\psi_2 = 0 \,.
$$
Now, if we make use of these two equations, then the following third order derivative 
$$ 
 D^3(\psi_i\psi_j)=\psi_i'''\psi_j+3\psi_i''\psi_j'+3\psi_i'\psi_j''+\psi_i\psi_j'''\,,
$$
can be rewritten as follows 
$$
 D^3(\psi_i\psi_j)=-(kv\,\psi_i'+kv'\,\psi_i)\psi_j-3kv\,\psi_i\psi_j'+3\psi_i'(-kv\,\psi_j)-
\psi_i(kv\,\psi_j'+kv'\,\psi_j),
$$
that after simplification it becomes 
$$ 
 D^3(\psi_i\psi_j)=-k[2v'\,\psi_i\psi_j+4v(\psi'_i\psi_j+\psi_i\psi'_j)] \,.
$$
We have therefore obtained 
$$ 
 D^3(\psi_i\psi_j)+4kv(\psi'_i\psi_j+\psi_i\psi'_j)+2k\,v'(\psi_i\psi_j) =0\,,
$$
what proves that the three functions $f_{ij}=\psi_i\psi_j$, $i,j=1,2$,  are  solutions of (\ref{Eqkv3prop1}).  
Finally, the Wronskian of these three functions  is given by 
$$ 
W[\psi_1^2, \psi_1\psi_2, \psi_2^2] = 2 (\psi_1\,\psi'_2-\psi_2\,\psi'_1)^3 \,. 
$$
Therefore, as $\{\psi_{1},\psi_{2}\}$ is a fundamental set of solutions of the second order equation (\ref{Eqkv2prop1}),  the functions $\psi_{1}^2$, $\psi_{2}^2$ and $\psi_1 \psi_2$ are linearly independent 
and they  span the linear space of  solutions of (\ref{Eqkv3prop1}).

\hfill$\Box$

%---------------------

\begin{prop}
\begin{enumerate}

\item Let $f$ be a solution of the projective vector field equation (\ref{pvfe}).  
Then the function $u$ such that $ ku = f' / f$ (where $k\ne 0$)
is a particular solution of the following second-order Riccati equation:
\beqn
 u'' + 3k\,u\,u' + k^2u^3 + 4kv\,u + 2kv' = 0\,. 
 \label{projReq} 
\eeqn

\item Suppose that $u_1(x)$ is a solution of the Riccati equation
\beqn
 \zeta' + k\zeta ^2 + kv = 0 \,. 
 \label{Eqzeta} 
\eeqn
Then the function $u=2u_1$ is a solution of the second-order Riccati 
equation (\ref{projReq}). 
\end{enumerate}
\end{prop}
{\it Proof:} 1. The equation  (\ref{pvfek}) is invariant under dilations, and
then we can choose
an adapted coordinate for which  dilations generator is $(1/k)\partial/\partial
z$ (i.e. we use a function $z$ instead of $y$ such that $y=e^{kz}$), and then
using the expressions analogous to (\ref{dilchange}) 
\begin{equation}
y'=kz'\, e^{kz}\,,\quad y''=(kz''+k^2z^{\prime 2})e^{kz} \,,\quad y'''=
(kz'''+3k^2z' z''+k^3z^{\prime 3})e^z\,,\label{dilchangek}
\end{equation}
the  differential  equation that we obtain does not depend on $z=y'/(ky)$ but on its derivatives, $z',z''$ and $z'''$,  and if we define $u=z'$, it becomes (\ref{projReq}), which is a generalization of the nonlinear oscillator equation. 
Here the coefficients are fixed by the projective vector field equation. 
In other words, the new reduced equation can be rewritten as
$$ 
 R^2(u)+4vR^0(u)+2v'=0.
$$

2. First, we note that the derivative of  $u'_1+ku_1^2+kv=0$ leads to the following second order equation  
\begin{equation}
u_1''+2ku_1u_1'+kv'=0  \,. 
\label{ddv1}
\end{equation}
Now putting $u=2u_1$ in the left hand side of (\ref{projReq})
we obtain 
$$  
u'' + 3ku\,u' + k^2u^3 + 4kv\,u + 2k\,v'=2u_1''+12ku_1\,u_1'+8k^2u_1^3+8ku_1\,v+2kv',
$$
and therefore taking into account (\ref{ddv1}):
$$
 u'' + 3ku\,u' + k^2u^3 + 4kv\,u + 2k\,v'=-2(2ku_1u_1'+kv')+12ku_1u_1'+8k^2u_1^3+8ku_1v+2k\,v'
$$
which symplifying terms reduces to 
$$
u'' + 3ku\,u' + k^2u^3 + 4kv\,u + 2k\,v' =8ku_1(u'_1+k\,u_1^2+k\,v)
$$
from where using (\ref{Eqzeta}) on the right hand side  we obtain the result of the Proposition.

This shows a relationship among solutions of  ordinary Riccati equations
and those of related second-order  Riccati equations.

%-----------------------------------------------
 
\subsection{Global projective vector field and integrable ODEs}

Since the solution space  of the projective vector field equation is spanned by
$$ 
\textrm{Span} ( \psi_{1}^2, \psi_{2}^2, \psi_1 \psi_2),
$$
an arbitrary solution of the projective vector 
field equation is given by
\beqn
\Psi = A\psi_{1}^2 + 2B\psi_1 \psi_2 + C\psi_{2}^2, 
\eeqn
an arbitrary linear combination of basis vectors. 
This is periodic when $\psi_1$ ad $\psi_2$ are periodic,  and hence it is a global solution of the equation projective vector field as a consequence of the existence of gobal solutions of the  Hill equation.
This $\Psi$ is called the {\it global projective vector field} \cite{Hit91}. 

Milne-Pinney equation can be written as a system of two first-order differential equations and it turns out to be a 
Lie system (see e.g. \cite{CLR08}). The same is true for the Hill equation (or equivalently for the harmonic oscillator with a time-dependent frequency).
As the Vessiot-Guldberg Lie algebra of both equations is the same we can determine a mixed superposition rule allowing to write the general solution of Pinney equation in terms of  two independent solutions of Hill equation (see \cite{CLR08}): 

%---------------------
 
\begin{prop}
If $\psi_1$ and $\psi_2$ satisfy Hill's equation -- i.e. they are periodic  solutions of 
\beqn
\frac{d^2\psi_i}{dx^2}+v(x)\psi_i=0,\quad i=1,2,\label{Hilleq}
\eeqn
 then the square root $\psi$ of the global projective vector field, that is, 
$\psi = \sqrt{  A\psi_{1}^2 + 2B\psi_1 \psi_2 + C\psi_{2}^2 }$  is a periodic function
satisfying the  Milne-Pinney equation  \cite{Mi30,Pinney50}
\begin{equation}
\psi^{\prime \prime} + v(x) \psi = \frac{\sigma}{\psi^3},   \label{MPeq}
\end{equation}
with $\sigma = AC - B^2$.
\end{prop}

Note also that we can consider the second order nonlinear differential equation \cite{CHM03}
\begin{equation}
f\, f^{\prime \prime}- \frac{1}{2}\,f^{\prime 2} +2v \, f^2
- \frac \sigma 2=0,\label{Gambier22}
\end{equation}
which with the change of variable $f=-\psi^2/2$ becomes (\ref{MPeq}). Moreover, taking derivative with respect to $x$ at equation (\ref{Gambier22}) we see that a solution of  such equation is a solution 
of the projective vector fiel equation  (\ref{pvfe}).

In a similar way, we can consider the second-order Kummer-Schwarz equation:
$$ 
\frac{1}{2}\frac{f^{\prime \prime}}{f} - \frac{3}{4}\left(\frac{f^{\prime}}{f}\right)^2 + \sigma f^2
+ v(x)=0,
$$
which is a particular case of the second-order Gambier equation \cite{CaGdL13} and it was recently analysed in 
\cite{LS13} from the perspective of Lie theory. 
It has been proved to be a Lie system associated with a Vessiot-Guldberg Lie algebra isomorphic to $\mathfrak{sl}(2,\mathbb{R})$, and therefore admitting a nonlinear superposition rule \cite{CadL11}. 
But as in the case of Milne-Pinney equation we can also find a mixed superposition rule in terms of solutions of Hill equation. 
More explicitly, the solution is given by
\beqn
f(x) = (A\psi_{1}^2 + 2B\psi_1 \psi_2 + C\psi_{2}^2)^{-1},
\eeqn
where $\psi_1$ and $\psi_2$ satisfy the Hill's equation and $B^2=AC-\sigma W^{-2}$, where $W$ is the Wronskian determinant of both solutions, $W=W[\psi_1,\psi_2]$ \cite{LS13}.

\bigskip

 It has also been proved in \cite{Reid71}  that if $\psi_1$ and $\psi_2$ satisfy Hill's equation, with $\psi_2(x_0)\ne 0$, and $W$ denotes the Wronskian, $W=W[\psi_1,\psi_2]$, then for  each nonzero real number $m$,  such that $0\ne m\ne1$,  the function 
$$ 
   \Psi = \left(\psi_{1}^{m} + \frac c{(m-1)W^2} \, \psi_{2}^{m}\right)^{1/m} 
$$ 
 satisfies the differential equation 
$$  
 y''+v(x)y = c \,\frac{(\psi_1\psi_2)^{m-2}}{y^{2m-1}}. 
$$
 There is a clear difference with respect to the preceding case of Milne--Pinney equation  because now the right-hand side of equation depends on the functions $\psi_1$ and $\psi_2$.

 This result was generalised in \cite{Reid73} where it is shown that if  $\psi_{1}$ and $\psi_{2}$ are functionally independent solutions of the linear
  homogeneous differential equation 
  $$
  y'' + r(x) \, y' +q(x)\, y=0, 
  $$
then the function 
$$ 
  \Psi = \left(A\psi_{1}^{m} + B\psi_{2}^{m}\right)^{1/m}
 $$ 
  is a solution of  the Reid type equation 
$$
y'' + r(x) \, y' +q(x)\, y=AB(m-1)(\psi_1\psi_2)^{m-2} \frac{W^2}{y^{2m-1}} ,
$$
where $W$ is as before the Wronskian of the two functions.

Moreover, if we set  $\psi = (\psi_1\psi_2)^{k/2}$ then $\psi$ satisfies Thomas equation \cite{Thomas52}
\beqn
y'' + r(x)y' +kq(x)y = (1-l) \frac{y^{\prime2}}{y} - \frac{1}{4}kW^2 y^{1- 4l},\qquad kl=1, 
\eeqn
where $W$ denotes the Wronskian of the two solutions.

We can recover reduced Gambier equation (or Painlev\'e-Gambier XXVII equation)\cite{Ince44} for special values
of $m$.

%-----------------------------------------------

\section{ Invariants, prolongations and  Riccati equation}

For a geometric approach to a partial differential equation in $m$ independent variables and one
dependent variable $v$, we must consider the space $M \times V =\{ (\mathbf{x}, u)\mid \mathbf{x}\in M, u\in V\}$,
where $M = {\Bbb R}^m$ and $V = {\Bbb R}$. Suppose that $G$ is a Lie group acting on some open subset $N \subseteq M \times V$. 
Then the transformation by $g \in G$ is
$$
g (\mathbf{x},u) = (\bar {\mathbf{x}}, {\bar u}),\qquad g\in G,$$
and a hypersurface $u=u(\mathbf{x})$ in $N$ is transformed into  another one, ${\bar u} = {\bar u}(\bar{\mathbf{x}})$.

An infinitesimal transformation is given by 
$$
\begin{array}{rcl}
\bar {\mathbf{x}} &=& \mathbf{x} + \epsilon\,\mathbf{f}(\mathbf{x},u) + \vartheta(\epsilon^2),\\
{\bar u} &=& u + \epsilon \,g(\mathbf{x},u) + \vartheta(\epsilon^2),
\end{array}
$$
wich can be understood as the flow of the
vector field in $M\times V$ 
\beqn
X = f^i(\mathbf{x},u)\frac{\partial}{\partial x^i} + g(\mathbf{x},u)\frac{\partial}{\partial u}.\label{inftrans}
\eeqn
 
We restrict ourselves to the case of  ordinary differential equations  (i.e.  $m=1$) and  then $M=\mathbb{R}$.
A (may be local) smooth section $\sigma$ for the projection $\pi:M \times V \to M$ defines 
 a smooth function $u= u(x)$, by means of  $\sigma(x)=(x,u(x))$ and then for each natural
 number $k$  induces a function 
$u^{(k)} = {\rm pr}^{(k)}u(x)$, called the $k$-th prolongation of $u$, where
$$
{\rm pr}^{(k)}u: \mathbb{R} \longrightarrow {\mathbb{R}}^{k+1}
$$
is the curve whose components are  the derivatives of $u$ of orders from $0$ to $k$.
The total space $M \times V^{(k)} \subseteq {\mathbb{R}}^{k+1}$, the coordinates
of which represent the independent variable ${x}$, the dependent variable $u$  and the derivatives
of $u$ to order $n$ is called the {\it $n$-th order jet space}
of the underlying space $M \times V$, sometimes denoted $J^n\pi$.

Similarly, a  vector field in $N\subset  M \times V$ given in local coordinates by
\beqn
X = f(x,u)\frac{\partial}{\partial x} + g(x,u)\frac{\partial}{\partial u}, 
\eeqn
admits  a $k$-order prolongation. We are only interested in the first-order 
prolongation \cite{Olver1995}. 
Note that if we consider the corresponding infinitesimal transformation (\ref{inftrans}), then
\beqn
\frac{d{\bar u}}{d{\bar x}} = (Dg - u_xDf) = g_x + (g_u - f_x)u_x
- f_uu_x^{ 2},
\eeqn
where $g_x=\partial g/\partial x$ and $g_u=\partial g/\partial u$, and similarly for $f_x$ and $f_u$, with $D$ being given by 
$$
D = \frac{\partial}{\partial x} + u_x\frac{\partial }{\partial u}.
$$
It  follows directly from
$$
\begin{array}{rcl}
{\displaystyle\frac{d{\bar u}}{d{\bar x}} }&=& {\displaystyle\frac{d(u + \epsilon g + 
\vartheta(\epsilon^2))}{d(x + \epsilon f + 
\vartheta(\epsilon^2))}
= \frac{u_x + (g_x + g_uu_x)\epsilon + \vartheta(\epsilon^2)}{1 + (f_x + f_uu_x)\epsilon 
+ \vartheta(\epsilon^2)}}\\
&&\\
&=& u_x+\epsilon \,(Dg - u_x Df) =u_x+ [g_x + (g_u- f_x)u_x
- f_uu_x^{2}]\epsilon + \vartheta(\epsilon^2).\end{array}
$$
This provides us with the following definition of first-order prolongation 
of  $X$\cite{Olver1995}:
%-------------------
\begin{dfn}
The prolongation ${\rm pr}^{(1)}(X)$ of the  vector field   $X\in\mathfrak{X}(N)$ is the vector field in $J^1\pi$ given by
\beqn
{\rm pr}^{(1)}(X)=X^{(1)} = f(x,u)\frac{\partial}{\partial x} + 
g(x,u)\frac{\partial}{\partial u} + 
(Dg - u_x Df)\frac{\partial}{\partial u_x}.
\eeqn
\end{dfn}
These prolongations will play a relevant r\^ole in the search for symmetries of differential equations as it is shown in
 next subsection where the first-order  Riccati equation is used as an example.

%-----------------------------------------------

\subsection{Infinitesimal symmetries of standard Riccati equation and projective vector field equation} 

The first-order differential equation for the function $u$
\begin{equation}
u_x=u^2+v(x),\label{SRE}
\end{equation}
where $v$ is now a given function, usually called standard Riccati equation,  is a particular case of the general Riccati equation:
\begin{equation}
u_x=a_2(x)u^2+a_1(x)u+a_0(x)\,,\label{GRE}
\end{equation}
 for $a_2=1$, $a_1=0$ and $a_0(x)=v(x)$.

The solutions of such non-autonomous differential equation (\ref{SRE}) are the integral 
curves of the vector field  in ${\mathbb{R}}^2$ given by
\begin{equation}
X=\pd{}x+(u^2+v(x))\pd{}u\,.\label{vfSRE}
\end{equation}

An infinitesimal Lie symmetry of such differential equation is represented by a vector field 
$$ 
Y=f(x)\pd{}{x}+g(x,u)\pd{}{u} 
$$
such that there exists a function $\tau $ satisfying   
\begin{equation}
[Y,X]=\tau \, X,\label{symcond1}
\end{equation}
because then the solutions of the differential equation are transformed under the flow of $Y$ into solutions, up to a reparametrization.

Taking into account that 
$$
[f\partial_x+g\partial_u,\partial_x+(u^2+v)\partial_u]=-(Xf)\partial_x+\left(fv_x+2gu-\partial_xg
-(u^2+v)\partial_ug\right)\partial_u,
$$
we find that the symmetry condition implies that $\tau=-f_x(x)$,  and that $f$ and $h$ are related as follows:
\begin{equation}
\partial_xg+(u^2+v)\partial_ug-fv_x-2gu=(u^2+v)\,f_x\,.\label{symcond2}
\end{equation}

We can  alternatively consider the given differential equation (\ref{SRE}) as  defining a $2$-dimensional 
submanifold in ${\mathbb{R}}^3$, with local coordinates $(x,v,w)$,
defined by  zero level set of the constant rank map $\phi:\mathbb{R}^3\to \mathbb{R}$, $\phi(x,u,w)=w-u^2-v(x)$, that is,
\beqn
\Sigma=\phi^{-1}(0)=\{(x,u,w)\mid w-u^2-v(x)=0\}.\label{SREdos}
\eeqn

The first-order   prolongation of $Y$ is given by 
\begin{equation}
Y^{(1)}=f(x)\pd{}{x}+g(x,u)\pd{}{u}+(g_x(x,u)-wf_x(x))\pd{}w\,
\end{equation}
and when we consider the  symmetry condition, which is but a tangency condition, $Y^{(1)}\phi_{|\Sigma}=0$, we find 
\begin{equation}
Y^{(1)}(w-(u^2+v(x)))_{|w=u^2+v(x)}=0\,,\label{symconddos}
\end{equation}
and more explicitly,
\begin{equation}
\partial_xg(x,u)+(w\partial_ug(x,u)-f_x(x)w)_{|w=u^2+v(x)}=f(x)v_x(x)+2g(x,u)\,u\,,
\end{equation}
which reproduces (\ref{symcond2}). 

We can now establish the following relationship among symmetries of the standard Riccati equation and solutions of the projective 
vector field equation.

%---------------------

\begin{prop}
The standard Riccati equation (\ref{SRE}) remains invariant with respect to the first prolongation 
${\rm pr}^{(1)}Y$ of the vector field 
$$ 
 Y=f(x)\pd{}{x}+g(x,u)\pd{}{u} 
$$
provided that $f$ satisfies the projective vector field equation.
\end{prop}

{\it Proof}: 
Once  that $f$ is fixed in (\ref{symcond2}) we can try to determine an affine function $g$ (i.e. of the form  $g(x,u)=k(x)u+h(x)$), satisfying such symmetry condition. Then the functions $k$ and $h$ must satisfy:
$$
k_x(x)u+h_x(x)+(u^2+v(x))(k(x)-f_x(x))=f(x)v_x(x)+2u(k(x)u+h(x)),
$$
and therefore
$$
u^2(k(x)+f_x(x))+u(2h(x)-k_x(x))+f(x)v_x(x)+v(x)(f_x(x)-k(x))-h_x(x)=0.
$$
This shows first that we must choose $k(x)=-f_x(x)$ -- i.e. 
the function $g$ is  such that
$$
g(x,u)=-f_x(x) \,u+h(x),
$$
and using such an expression for $g(x,u)$ in the symmetry condition  (\ref{symcond2}) we find 
$$
u(f_{xx}(x)+2h)+f(x)\,v_x(x)+2v(x)f_x(x)-h_x(x)=0,
$$
which shows that we should choose $h$ 
such that $h=-\frac 12 f_{xx}(x)$, and then replacing $h$ by this value in the
preceding equation we find that $f$ must be such that
$$
\frac 12 f_{xxx}(x)+2v(x)f_{x}(x)+f\,v_x(x)=0\,,
$$ 
and hence as indicated in \cite{Hit91} $f$ is a solution of the projective vector field equation (\ref{pvfe}). 

%-------------------
{\bf Remark}: 
One would obtain the same result if one starts from the differential equation 
$u' = a_2(x)u^2 + a_1(x)u + a_0(x)$. In this case $u$ must
be expressed in terms of $a_2$, $a_1$ and $a_0$ and their derivatives.

%-----------------------------------------------
 
\subsection{Integrals of motion  and other dynamical features}

A constant of the motion or symply an  integral of the motion for a system of ODE's
\beqn\label{eom} 
\frac{dy_i}{dx}=X_i(x,y_1,\cdots,y_n)\hskip 10pt i=1,\cdots,n\eeqn
 is a non-constant differentiable function
 $\Phi(x,y_1,\cdots,y_n)$ that retains a constant value on any  integral curve of the system.  
 This means  its derivative with respect  to $x$ vanishes on the solution curves:
 \beqn\label{iom} 
 \frac{d\Phi}{dx}=0\Rightarrow \frac{\partial\Phi}{\partial x}+\sum_i\frac{\partial\Phi}{\partial
 y_i}\frac{dy_i}{dx}=0\Longrightarrow  \widetilde{D}[\Phi]=0, 
 \eeqn
 where 
 $$\widetilde{D}:=\frac{\partial}{\partial x}+\sum_i X_i\frac {\partial}{\partial
 y_i} 
 $$ 
 is called the material or total  derivative. For an autonomous system this reduces to
\beqn
D\Phi=\sum_iX_i\frac{\partial\Phi}{\partial y_i}=0 
\eeqn
 where  
 $$ 
 D=\sum_iX_i\frac{\partial}{\partial y_i} 
 $$  
 is just the vector  field associated with the given autonomous system. 
 The $x$-independent integrals of motion are usually called first-integrals.
 
 Higher order differential equations can be written as an associated system of first-order differential equations and then the constants of motion for such system are called constants of motion for the higher-order differential equation.
 As an instance  the projective vector field equation (\ref{pvfe}) can be written as the system
 $$\left\{\begin{array}{rcl}{\displaystyle \frac{dy}{dx}}&=&w\\{\displaystyle \frac{dw}{dx}}&=&a\\
 {\displaystyle \frac{da}{dx}}&=&-4vw-2v'y
 \end{array}\right.
 $$
 and then  the function $\Phi(x,w,a)=2v(x) y^2-\frac 12 w^2+ya$ is a constant of the  motion, because the vector field 
 $$ 
  X=\pd{}x+w\pd{}y+a\pd {}w-(4vw+2v'y)\pd{}a 
$$ 
 is such that $X\Phi=0$. 
 The constant of motion can also be rewritten as 
$$
\Phi(x,y,y',y'')=2v(x)y^2-\frac 12 y^{\prime 2}+y\, y''
$$
which is a constant of the motion for the projective vector field equation (\ref{pvfe}).  This shows that a nowhere vanishing  function is solution of 
 the projective vector field equation (\ref{pvfe}) if and only if there exist a constant $C$ such that the function is solution of the 
 second-order  differential equation 
 $$y''=2v(x)\, y -\frac {2y} y^{\prime 2}.
 $$

With an analogous procedure  one can compute the first-integrals of similar type for a third-order differential
equation. For example, the first-integral of the stationary Calogero-Degasperis-Ibragimov-Shabat equation \cite{EEL11}.
\beqn\label{CDIS} y''''+3y^2\, y''+9y\, y^{\prime 2}+3y^4 y' =0 
\eeqn
has associated a vector field 
$$ 
 X=w\pd{}y+a\pd {}w-(3y^2a+9yw^2+3y^4w)\pd{}a 
 $$
and  the function 
$$\Phi(y,w,a)= 2ay+6y^3w+y^6-w^2 
$$ 
 is such that $X\Phi=0$  and therefore is  constant of motion.  
The corresponding function 
 given by 
$$ 
 \Phi(y,y',y'')= y\bigl(2y''+ (6yy' + y^4)y\bigr) -y^{\prime 2}
$$
$$ 
 f\bigl(2f_{xx} + (6ff_x + f^4)f\bigr) - {f_x}^2, 
 $$
 is a particular case of the previous one with  $v = (3/2) y y' + (1/4) y^4$. The remarkable point here is that one can interpret equation (\ref{CDIS}) in terms of the stabilizer set of the 
coadjoint action. The function $y=f(x)$ is a solution of (\ref{CDIS}) if and only if 
$$  
 \textrm{ad\,}_{f\frac{d}{dx}}^{\ast} (3ff_x + (1/2) f^4) dx^2 = 0. 
$$

%-----------------------------------------------
 
\section{Second-order Riccati equation, Painlev\'e II and higher Painlev\'e type equations}

Consider once again the linear third-order equation (\ref{pvfe}) 
and the corresponding second-order Riccati equation (\ref{projReqk1}).

Let us define   for each function $f$ the functions
\beqn
u_1 = \frac{f'}{f}, \qquad u_2 = \frac{f''}{f} - \frac{1}{2}\left(\frac{f'}{f}\right)^2 + w, \label{defues}
\eeqn  
where $w$ is an arbitrary  but fixed function. Then,
\begin{equation}
u'_1=\frac d{dx}\left( \frac{f'}{f}\right)= \frac{f''}{f}-\left(\frac{f'}{f}\right)^2=u_2 - \dfrac{1}{2}u_{1}^{2} - w  .\
\end{equation}
Moreover,
$$ 
 u'_2=\frac{f'''}{f}-\frac{f''\,f'}{f^2}-u_1u'_1+w'
$$
and if we use the second equation in  (\ref{defues}) to write 
$$ 
 \frac{f''}{f}=u_2+\frac 12 u_1^2-w,
$$
we see that when $f$ satisfies equation 
(\ref{pvfe}) and therefore 
$$  
\frac{f'''}{f}=-4v\,u_1-2v',
$$
then 
$$u'_2= -4v\,u_1-2v'-u_1\left(u_2+\frac 12 u_1^2-w\right)-u_1u'_1+w'$$
that is, 
$$
u'_2=-4v\,u_1-2v'+u_1(w-u_2)-u_1\left(u'_1+\dfrac{1}{2}u_{1}^{2} \right)+w'.
$$
Consequently, such functions  $u_1$ and $u_2$ satisfy the system of differential equations
\begin{equation}
\left\{
\begin{array}{rcl}
u_1' &=& u_2 - \dfrac{1}{2}u_{1}^{2} - w \,, \\ &&\\
u_2' &=& 2(w - 2v)u_1 - 2u_1u_2 - 2v' + w'.
\end{array}\right.\label{systemues}
\end{equation}
Conversely, if  $u_1$ and $u_2$  are solution of this system then 
$f(x)=\exp\left(\int^xu_1(\zeta)\ d\zeta\right)$ is a solution of (\ref{pvfe}), because, by definition of $f$, 
$$\frac{f'}{f}=u_1, \quad \frac{f''}{f}=\frac d{dx}\left(\frac{f'}{f}\right)+\left(\frac{f'}{f}\right)^2=u'_1+u_1^2= u_2 + \dfrac{1}{2}u_{1}^{2} - w,
$$ 
and 
$$
\frac{f'''}{f}=\frac d{dx}\left(\frac{f''}{f}\right)+\frac{f''\,f'}{f^2}=u'_2+u_1u_1'-w'+u_1\left(u_2 + \dfrac{1}{2}u_{1}^{2} - w\right)=u'_2-w'+2u_1(u_2  - w),
$$
from here we see that 
$$ 
\frac{ f^{\prime \prime \prime} +4v\,f^{\prime} + 2v^{\prime}\,f }f=u'_2-w'+2u_1(u_2  - w)+4v\,u_1+2v'
$$
and the second condition (\ref{systemues}) shows that $f$ is  solution of  (\ref{pvfe})

The very essence of the second-order Riccati equation  (\ref{projReqk1}) is that it is obtained as a system of two
coupled differential equations. One starts from a first-order Riccati equation $u_1' = u_2 - \frac{1}{2}u_{1}^{2} - w$, 
where $w$ is an arbitrary differentiable  function of the independent variable $x$ and then couple $u_1$ to
$u_2$ through a  linear  differential equation in $u_2$ involving $u_1$.

We wish to express these two equations in terms of $u_2$. It is clear from the second equation that
$$
u_1 = \frac{1}{2}\frac{u_2' + 2v' - w'}{w - 2v -u_2} 
$$ and if we define a new function $K(x)$ by 
$K=\pm (u_{2} -w +2v) $ we can write 
$$u_1 = - \frac{1}{2}\frac{K'}{K},
$$ and if we reexpress the first equation of (\ref{systemues}) in terms of $K$ 
and  substitute $u_2$ by $w-2u\pm K$  we obtain 
\beqn
K'' = \frac{5}{4}\frac{K^{\prime 2}}{K} + 4vK \mp 2K^2.
\eeqn
We can put  it into  a more amenable form if we substitute $K = - 1/p$, and now we obtain
\beqn
p'' = \frac{3}{4}\frac{p^{\prime 2}}{p} - 4vp \mp 2.
\eeqn 
This is a reduced Gambier equation or G5 equation in Gambier's
classification \cite{Gam}.

\medskip

%-------------------
{\bf Remark} 
(a) \, If we assume $v$ to be a constant such that $v = \frac{\alpha}{4}$, 
then the second-order Riccati equation (\ref{projReqk1})
boils down to the modified Emden equation
\beqn
u'' + 3u\,u' + u^3 + \alpha u = 0,
\eeqn
which therefore can be obtained from the pair of differential  equations
$$\left\{\begin{array}{rcl}
u_1' &=& u_2 - \frac{1}{2}u_{1}^{2} - w, \\
u_2' &=& (2w - \alpha)u_1 - 2u_1u_2 + w'.
\end{array}\right.
$$

\medskip

(b) \, Suppose we take $u = 2\Psi' / \Psi$. Then, 
the  second-order (projective) Riccati 
equation (\ref{projReqk1}) can be related to  the family of  Milne--Pinney equations \cite{Mi30,Pinney50}
\beqn
\Psi'' + v\Psi = \frac{\sigma}{\Psi^3},
\eeqn
where $\sigma$ is some constant. In fact, such a family is described by the third order differential equation 
$$ (\Psi^3\Psi^{\prime\prime} + v\Psi^4)^{\prime} = 0.$$ 

This is a differential equation invariant under dilations and Lie recipe for reduction amounts to define  a new variable $ \zeta$ such that $\Psi=e^{\frac 12 \zeta}$ (i.e. $u=\dot \zeta = 2\Psi' / \Psi$), 
and then replacing
$$
\frac{\Psi'}{\Psi}=\frac 12 u,\quad \frac{\Psi''}{\Psi}=\frac 12 u'+\frac 14 u^2,\quad \frac{\Psi'''}{\Psi}=\frac 12 u''+\frac 34 uu'+\frac 18 u^3,\quad 
$$
in the equation 
$$\Psi^3\Psi^{\prime \prime \prime} + 3\Psi^2\Psi^{\prime}\Psi^{\prime \prime} + 4v\Psi^3\Psi^{\prime} + v^{\prime}\Psi^4 = 0,
$$
 we obtain  the second-order differential equation (\ref{projReqk1}). 
This immediately yields our result.
$\Box$

\bigskip

The projective vector field equation (\ref{pvfe}) admits a Lax formulation in the following sense. 
Given a function $f$ define a pair of matrices $P$ and $Q$ as follows:
$$
P=\begin{pmatrix} f'/2&-\int^xv(\zeta)f'(\zeta)\,d\zeta \\-f&-f'/2\end{pmatrix},\qquad Q=\begin{pmatrix}0 & v  \\
-1 & 0  \end{pmatrix},
$$
for which
$$QP=\begin{pmatrix} -vf&-vf'/2\\ -f'/2&\int^xv(\zeta)f'(\zeta)\,d\zeta  \end{pmatrix}, \qquad PQ=\begin{pmatrix}\int^xv(\zeta)f'(\zeta)\,d\zeta &vf'/2\\f'/2&-vf  \end{pmatrix}.
$$

Therefore, the matrices $P$ and $Q$ are a Lax pair,  that is, 
\beqn\label{loopcoad}
P' + [Q,P] = 0,
\eeqn
if and only if $f$ is a solution of the differential equation 
$$
f''+2v\,f+2\int^xv(\zeta)f'(\zeta)\,d\zeta =0,
$$
and taking derivatives we find 
 the projective vector field equation (\ref{pvfe}). 

The corresponding Lax formulation for  the  projective second-order Riccati equation (\ref{projReqk1}) will be given by the same matrix $Q$ and
$$P=\begin{pmatrix}\frac 12 ue^{\int^xu(\zeta)\, d\zeta} &-\int^xv(\zeta)u(\zeta)e^{\int^\zeta u(\zeta')\, d\zeta'} \,d\zeta \\- e^{\int^xu(\zeta)\, d\zeta} &-\frac 12ue^{\int^xu(\zeta)\, d\zeta} \end{pmatrix}, 
$$
and then the matrices $P$ and $Q$ are a Lax pair if and only if $u$ is a solution of the integro-differential equation 
$$
(u'+u^2+2v)e^{\int^xu(\zeta)\, d\zeta} +2\int^xv(\zeta)u(\zeta)e^{\int^\zeta u(\zeta')\, d\zeta'} \,d\zeta=0.
$$

Now, taking derivative with respect to $x$ we find 
$$(u''+2uu'+2v')e^{\int^xu(\zeta)\, d\zeta} +u(u' +u^2+2v)e^{\int^xu(\zeta)\, d\zeta} +2uve^{\int^xu(\zeta)\, d\zeta}=0,
$$
and simplifying the common factor we obtain the second-order Riccati equation (\ref{projReqk1}).

{\bf Remark} The equation (\ref{loopcoad}) has a nice geometric interpretation in terms of the coadjoint action of loop 
algebra \cite{PSbook,RSTS}. The loop group $C^{\infty}(S^1,G)$ is the group of smooth functions 
on the circle $S^1$ with values on
real semi-simple Lie group $G$. Its Lie algebra $C^{\infty}(S^1,{\frak g})$ has a central extension provided by 
the Kac-Moody cocycle $\omega(A,B) = \int_{S^1} <A(x),\frac{dB(x)}{dx}> dx$. The coadjoint representation of the extended 
loop algebra is given by 
$$ad^{\ast}(P,a)(Q,b) = \Big([P,Q] + b\frac{dP(x)}{dx}, 0 \Big), $$
for any $P,Q \in C^{\infty}(S^1,{\frak g})$ and $a,b \in {\Bbb R}$. The equation (\ref{loopcoad}) is the {\it stabilizer set} of the coadjoint orbit confined to hyperplane $b = -1$. Thus once again we obtain equation which can be 
interpreted in terms of stabilizer orbit.

%-----------------------------------------------

\subsection{Painlev\'e II, B\"acklund transformation and second-order Riccati equation}

We wish to discuss integrable class of Painlev\'e II equation; 
in other words, we are interested in the rational
solutions for integer valued parameter $\alpha$ of the Painlev\'e II equation and their explicit characterization in terms of the Airy function.  
It is worth to note that half-integer valued parameter of the Riccati equation are also characterized by the Airy function.
 Here the Painlev\'e II equation (PII) is an ordinary nonlinear  second-order
differential equation with a parameter $\alpha$,
\beqn u'' = 2u^3 + xu + \alpha.  \label{PainleveIIb}
\eeqn
This equation has exactly one rational solution for $\alpha$ being an arbitrary
integer and has no rational solution if $\alpha$ is not an integer \cite{Clarks06,GramaRamani,Gromak}.
It admits a B\"acklund transformation $(u, \alpha) \mapsto (u,-\alpha)$, which clearly maps
rational solutions into rational solutions.

Consider now the Airy differential equation 
\beqn
\psi''+ x\psi = 0.
\eeqn
It is clear that the Airy differential equation is a particular instance of the Hill's equation (\ref{Hilleq})  for the special choice  $v(x) = x$.

%---------------------

\begin{prop}
1.- If $\psi$ is a solution of  the Airy equation
$$
\psi^{\prime \prime} + x\psi = 0,
$$
then the  function $u_1 = {\psi'}/{\psi}$ satisfies the  Riccati equation $u'+u^2+x=0$.

2.- The corresponding second-order Riccati equation  for the function $u=\lambda \,u_1$  is the 
Painlev\'e II equation
\beqn
u^{\prime \prime} = \frac 2{\lambda^2} u^3 + 2x\,u - \lambda, \label{sore}
\eeqn
which after a rescaling becomes the 
Painlev\'e II equation

\end{prop}

{\it Proof}: 1.- It suffices to use the expression (\ref{genhoRiceq}) for reduction recipe with $n=2$, $a_0(x)=x$ and  $a_1(x)=1$
(i.e. $R^1(u)+x\, R^0(u)=0$),  and expressions (\ref{firstRiceqs}),  
then $u_1 = {\psi'}/{\psi}$ is a solution of 
$$ 
  u'+u^2+x=0.
$$

2. Note that the function  $u=\lambda \,u_1$  satisfies the differential equation  $u'=-\frac {u^2}\lambda-\lambda \, x$ and therefore, deriving with respect to $x$ we find that the second-order Riccati differential equation satisfied by the function $u=\lambda \,u_1$ is 
$$u''=-\frac {2u}\lambda\left(-\frac {u^2}\lambda-\lambda \, x\right)-\lambda=\frac {2u^3}{\lambda^2}+2x\, u-\lambda.$$
If we now consider a new independent variable $\bar x=\mu \, x$ with $\mu\ne 0$, the  differential  equation is transformed into 
$$\frac{d^2u}{d\bar x^2}= \frac {2u^3}{\mu^2\lambda^2}+2\frac{\bar x}{\mu^3}-\frac \lambda{\mu^2},$$
and by choosing the parameters $\lambda$ and $\mu$ as $\mu^3=2$ and $\lambda=\mu^{-1}$ we see that the differential equation becomes a Painlev\'e II type differential equation:
$$\frac{d^2u}{d\bar x^2}= 2u^3+\bar x\, u-\frac 12. 
$$
This proves that the second order Riccati equation can be brought back to `standard Painlev\'e II' by an appropriate scaling.
$\Box$

Under the map $x \to -x$, it takes 
the form of a similar equation but  with  the opposite sign, that is, i.e. if $\psi$ is a solution of 
 the Airy differential equation 
$$
\psi'' - x\psi = 0,
$$
then $u_1 = \lambda\,   {\psi'}/{\psi}$ satisfies the differential  equation
\beqn
u^{\prime \prime} = \frac 2{\lambda^2} u^3 -2x\,u - \lambda
\eeqn
that can be transformed into a Painlev\'e II equation with an appropriate change of independent variable as before.

\smallskip

\noindent
%-------------------
{\bf Remark} ~The Painlev\'e transcendents (P-II -- P-VI) possess 
{\it B\"acklund transformations}
which map solutions of a given Painlev\'e equation into solutions
of the same Painlev\'e equation, but with different values of the
parameters. Therefore two Painlev\'e equations for $\alpha = 2$
and $\alpha = -2$ are connected by B\"acklund transformations.

\smallskip

\noindent
%-------------------
{\bf Remark} 
Considering $\psi'' \pm (x/2) \psi =0$ we can also relate Painlev\'e II equations
for parameters  $\alpha = \mp2$ and these are connected by the  B\"acklund transformations.

\smallskip

\noindent
%-------------------
{\bf Remark} \  If $y=f(x) $ is a solution of the differential equation (\ref{pvfe}), then for a given real number $\beta$ the function 
$y=f(x) $ is a solution of the differential equation 
\beqn  
 y^{\prime \prime \prime} + 4\bar v\,y^{\prime} + 2\bar v^{\prime}\,y +6\beta y\, y'=0 
 \end{equation}
 with $\bar v=v-\beta \, f$. Now if the  function $ f$ is positive and define the positive function $\psi$ by $f=\psi^2$, then as
  $$f'=2\psi\,\psi' ,\qquad
   f''=2\psi^{\prime 2}+2\psi\,\psi'',\qquad
    f'''=6\psi'\,\psi''+2\psi\, \psi''',
  $$
   we see that $\psi$ is then a solution of 
   $$2y\, y'''+6y'\, y''+8 \bar v\,y\,y'+2\bar v'\,y^2+12\beta\,y^3y'=0,
   $$
   which can be rewritten as 
$$(y''\, y^3+\bar v\,y^4+\beta \, y^6)'=0,
$$
i.e. there exists a constant $\sigma$ such that $\psi$ is a solution of the differential equation 
$$
y^{\prime \prime} + \bar v\, y + \beta \,y^3 = \frac{\sigma}{y^3}.
$$
For the particular case  $ \bar v = x/2$, that means that $\psi$ is a solution of the Ermakov-Painlev\'e II equation. Conte showed how one can transform it to
Painlev\'e II (see e.g. \cite{CR14})

\smallskip

\noindent
\paragraph{Remark}
Let us briefly describe the connection between Painlev\'e II hierarchy
and our approach. It is readily clear that the projective vector field
equation is the stability algebra of Virasoro orbit. In other words, instead of considering the symplectic structure defined by 
${\cal O}_1 = \partial_{x}$ and the Hamiltonian function $\frac 1{3!}u^3-\frac 12 u_x^2$, i.e.
$$u_t=\pd{}x\left(
\frac \delta{\delta u}\left[\frac 1{3!}u^3-\frac 12 u_x^2\right]
\right)=\pd{}x\left(\frac 12 u^2+u_{xx}\right)=u\, u_x+u_{xxx},$$
we consider
the second Hamiltonian structure of the KdV equation
\beqn
{\cal O}_2 = \partial_{x}^{3} + \frac 23 u\partial_x + \frac 13 u_x.
\eeqn
together with the Hamiltonian density $\frac 12 u^2$.
Using ``frozen Lie-Poisson structure'' we can define the first
Hamiltonian structure of the KdV equation too. This satisfies
famous Lenard scheme
\beqn
\partial_x {\mathcal{H}}_{n+1} = \left(\partial_{x}^{3} +  \frac 23 u\partial_x + \frac 13 u_x\right){\mathcal{H}}_{n},
\eeqn
where 
$$
{\mathcal{H}}_{1}= u, \qquad  {\mathcal{H}}_{2} = \frac{u^2}{2}, \qquad {\mathcal{H}}_{3} 
= \frac{1}{3!}u^3 - \frac{1}{2} u_{x}^{2}, \qquad \cdots
$$
are the conserved densities.
The mKdV hierarchy is obtained from the KdV hierarchy through Miura map
$u = v_x - v^2$. 
The second  Painlev\'e hierarchy is given
recursively by N. Joshi from the modified KdV hierarchy \cite{Joshi04} 
\beqn
P_{II}^{n} (u, \beta_n) \equiv \left(\frac{d}{dx} + 2v\right){\mathcal{J}}_n(u_x - u^2) 
- x\,u - \beta_n = 0,
\eeqn
where $\beta_n$ are constants and ${\cal J}_n$ is the operator defined by the first and second
Hamiltonian structures of the KdV equation
$$
\partial_x{\cal J}_{n+1}(u) = (\partial_{x}^{3} + 4u\partial_x + 2u_x){\cal J}_{n}(u)
$$
with ${\cal J}_1(u) = u$.

%-------------------
\begin{lem}
Let $\psi_1$ and $\psi_2$ be two linearly independent solutions of Hill's equation. 
Then, 

(a) The differential equations 
\beqn
 y^{(iv}  + 10vy^{\prime\prime} 
+ 10v^{\prime}y^{\prime}
+ (9v^2 + 3v^{\prime\prime})y = 0
\eeqn
trace out a four-dimensional space of solutions spanned by
$\{\psi_{1}^{3},\, \psi_{1}^{2}\psi_2,\, \psi_{1}\psi_{2}^{2},\, \psi_{2}^{3}\}.
$ 

(b) The differential equations 
\beqn
 y^{(v}  + 20vy^{\prime\prime\prime} 
+ 30v^{\prime}y^{\prime\prime} + 18v^{\prime\prime}y^{\prime} + 64v^2y^{\prime} + 4v^{\prime\prime\prime}y+
+ 64vv^{\prime}y = 0
\eeqn
traces out a five-dimensional space of solutions spanned by
$\{\psi_{1}^{4},\, \psi_{1}^{3}\psi_2,\, \psi_{1}^{2}\psi_{2}^{2},\, \psi_{1}\psi_{2}^{3},\, \psi_{2}^{4} \}.
$ 
\end{lem}

{\it Proof}: By a direct lengthy computation.
$\Box$

\bigskip

Using the standard projective transformation $u = y'/y$ we obtain 
the third-order and fourth-order Riccati equations associated to these equations, respectively, which  are, according to (\ref{moreRiceqs})  given by
\beqn
u''' + 4u\,u'' + 3u^{\prime 2} + 6u^2\,u' + 10v\,u'
+ u^4 +  10v\,u^2 + 10v'\,u + 9v^2 + 3v'' = 0 \label{thirdorderR1}
\eeqn
and 
$$
\begin{array}{rcl}
u^{(iv} + 5u\,v''' &+& 10u'u'' + 15uu^{\prime 2} + 10u^2u'' + 10u^3u' + u^5 
+ 20v(u'' + 3uu' + u^3)\\
&+& 30v'(u' + u^2) + 18v''u + 64v^2u + 4v''' + 64vv' = 0.
\end{array}
$$

Note that if we put $v =0$ then, all these equations form a Riccati hierarchy. 

\subsection{Higher Riccati equations and higher-order Painlev\'e class systems}

One can easily check that the third-order Riccati equation (\ref{thirdorderR1}) can be transformed  by setting $v = u^2$ into
a special case of the Chazy equation $XII$, given by
\beqn
u''' + 10uu'' + 9u^{\prime 2} + 36u^2u'
+ 20u^4  = 0.
\eeqn

Recently Ablowitz {\sl et al} \,\cite{ACH1,ACH2}  studied a general class of Chazy equation, defined
as
\beqn 
u''' - 2uu'' + 3u^{\prime2} = \alpha(6u' - u^2)^2.
\eeqn
where $\alpha$ is a real number.  The particular case $\alpha=1/16$ was mentioned in \cite{Cos2} and that of $\alpha= \frac{4}{36 - n^2}$, i.e.  
\beqn 
u''' - 2uu'' + 3u^{\prime2} = \frac{4}{36 - n^2}(6u' - u^2)^2.\label{chazyXII}
\eeqn
 has a singled value general solution.  More explicitly, 
this equation was first written down and solved by Chazy \cite{Chazy1,Chazy2} and is
known today as the generalized Chazy equation. Clarkson and Olver showed
that a necessary condition for the equation (52) to possess the Painlev\'e
property is that the coefficient must be $\alpha = \frac{4}{36 - n^2}$
with $ 1 < n \in {\Bbb N}$, provided that $ n \neq 6$. It has been
further shown in \cite{ClarkOlver} that the cases $n = 2,3,4$ and $5$,correspond to the dihedral
triangle, tetrahedral, octahedral and icosahedral symmetry classes.

\smallskip

\begin{prop}
If $\bar u$ is a solution of the third Riccati equation 
\begin{equation}
u''' + 3u^{\prime 2} + 4uu'' + 6u^2u' +u^4 = 0, \label{thirdorderR2}
\end{equation}
 then, using $\bar x=-x$ as independent variable, the function   $u=2\,u_1$ satisfies 
\beqn
u''' - 2uu'' + 3u^{\prime 2} = \frac{1}{8}(6u'- u^2)^2.\label{thirdorderR3}
\eeqn
\end{prop}

{\it Proof}: In fact, if $\bar u$ is a solution (\ref{thirdorderR2}) then $u=2\, \bar u$ is solution of 
$$
u'''+\frac 32  v^{\prime 2} +2u\, u'+\frac 32 u^2\, u'+\frac 18 u^4=0,
$$
and with the mentioned change of independent variable,
$$
u'''-\frac 32  v^{\prime 2} -2u\, u'+\frac 32 u^2\, u'-\frac 18 u^4=0,
$$
which can be rewritten as in (\ref{thirdorderR3}).
$\Box$ 

\bigskip

\noindent{\bf Remark} 
The Chazy IV equation
\beqn
u''' = -3uu'' - 3u^{\prime 2} - 3u^2 u'
\eeqn
is a derivative of second-order Riccati equation or second member of the Riccati chain gven in (\ref{firstRiceqs}) with $k=1$.

\bigskip

We are able to construct fourth-order equations of the Painlev\'e class family, derived by Bureau \cite{Bu}.
The Painlev\'e classification of the class of differential equations of the $4$-th order first and second degree 
was studied by Cosgrove \cite{Cos1,Cos2}. The subcase which will be relevant here is the Bureau symbol P1.

Cosgrove presented the results of the Painlev\'e classification 
for fourth-order differential equations where the Bureau symbol is P1. He gave
a long list of the equations F-VII -- F-XVIII in this category. Six equations, 
denoted by F-I, F-II, ..., F-VI, have Bureau
symbol P2. It is worth to note that all the cases with symbols P3 and P4 were found 
to violate a standard Painlev\'e test,
they admit non-integer resonances.  

\bigskip

We derive the following different equations from the list of Cosgrove on
the Painlev\'e classification of fourth order equations with Bureau symbol P1.

\begin{prop}
The following two equations follow from the higher Riccati equations
\begin{equation}
\begin{array}{lrcl}
\textrm{F-XII}&  \qquad u^{(iv} &=& - 4uu''' - 6u^2u'' - 4u^3u' 
- 12 uu^{\prime 2} - 10u'u'' \\
\textrm{F-XVI}& \qquad u^{(iv} &=& -5uu''' - 10u'uv'' - 15 uu^{\prime2} 
- 10u^2u'' 
- 10u^3u' - u^5  \\
 &&+& A(x)(u''' + 4uu'' + 3u^{\prime 2} + 6u^2u'+ u^4)
+ B(x)(u'' + 3uu' + u^3(x))\nonumber \\ &&+& C(x)(u^2 + u')
+ D(x)u + E(x) = 0. \,
\end{array}
\end{equation}
\end{prop} 

{\it Proof}: A) The F-XII fourth order equations with Bureau symbol P1 follows directly from the expressions  (\ref{moreRiceqs}) for the  fourth order Riccati $R^4(u)=0$
and the third-order Riccati $R^3(u)=0$ equations
 (also known
as Burgers higher-order flows) by means of the relation 
$$ 
\textrm{F-XII } :  R^4 (u)- u\,R^3(u)=0.
$$

B) The  F-XVI fourth order equations with Bureau symbol P1 is the combination
of all higher order Riccati equations.
$\Box$

\subsection{Second degree Painlev\'e II equation}

We have seen in the earlier section how Painlev\'e II is connected to second-order  Riccati or
projective vector field equation. In this section we carry out this investigation further to incorporate
the second degree Painlev\'e II equation.

The Hamiltonian of the standard Painlev\'e II equation $u'' = 2u^3 +x\,u + \alpha$ is given by
\beqn
H(x,u,w) = \frac{w^2}{2} - \left(u^2 + \frac{x}{2} \right)w - \left(\alpha + \frac{1}{2}\right)u,
\eeqn
where $u$ and $w$ stand for coordinate and momentum. The Hamiltonian equations of motion yield a set of Riccati equations
\beqn
\left\{
\begin{array}{rcl}
u^{\prime}& =& \dfrac{\partial H}{\partial w} = w - u^2 - \dfrac{x}{2}, \\
w^{\prime} &=&- \dfrac{\partial H}{\partial u} =
2u\,w   +\alpha +\dfrac{1}{2}      
\end{array} \right.
\label{systHPII}                                                                                                     .
\eeqn
This system was studied by Morales \cite{M06} who proved that for $\alpha\in \mathbb{Z}$ the system is not integrable by means of rational first integrals.

Taking derivative with respect to $x$ at the first equation,  using for $w$ the value obtained form it, i.e. $w=u^{\prime} +u^2 +\frac{x}{2}$, and replacing $w'$ by the value given by the second equation we obtain
$$u''=2u\left(u'+u^2 + \frac{x}{2}\right) +\alpha + \frac{1}{2}-2u\,u'-\frac 12= 2u^3+u\,x+\alpha,
$$
and therefore $u$ satisfies the  Painlev\'e II type differential equation (\ref{PainleveIIb}).

Now, if $u(x)$ and $w(x)$ are solutions of the system (\ref{systHPII}), the function $h(x)$ defined by 
$h(x)=H(x,u(x),v(x))$ is such that 
$$h'(x)=\pd Hx(x,u(x),v(x))=-\frac w2,\qquad 
h''(x)= -\frac {w'}2=-\frac 12 \left(2u\,w+\alpha+\frac 12\right)
$$ 
and then one easily check that 
\beqn
\left(h''(x)\right)^{2} + 4\left(h'(x)\right)^{3} + 2\,h'(x) \left(x\,h'(x) - h(x)\right) - \frac 14\left(\alpha + \frac{1}{2}\right)^2 = 0.\label{eqforh}
\eeqn

Conversely, if a function $h(x)$ satisfies the relation (\ref{eqforh}),  then the functions 
$$u(x)=\frac 12 \frac{h''(x)}{h'(x}+\left(\alpha +\dfrac{1}{2} \right)\frac 1{4h'(x)}, \qquad w(x)= -2\,h'(x)$$
are solutions of the system  (\ref{systHPII}).

Observe that the right hand side of the expression 
\beqn 
-2\,h' = w = u^{\prime} + u^2 + \frac{x}{2}.
\eeqn 
is the term appearing from the reduction of the second-order linear differential equation $y''+ \frac{x}{2}y=0$
when considering Lie recipe of invariance under dilations, i.e. $u = y'/y$. Therefore,
Hill's equation
\beqn y'' + \left(2\,h'(x) + \frac{x}{2}\right)y = 0, \eeqn
  leads by reduction from dilation symmetry to the preceding equation.

Note that the solution $u$ can be found directly from the function $h$ the expression using the second expression in (\ref{systHPII}):
\beqn
u =  \frac{2h''+ \alpha + \dfrac{x}{2}}{4h'}.
\eeqn

Thus we can extract several important information about the Painlev\'e II equation from the second-order 
projective Riccati of projective vector field equation.

%-----------------------------------------------

\section{Outlook}

The Riccati equation plays a very important role in mathematical physics and dynamical systems. 
The second and higher-order Riccati equations also play a big role in integrable ODEs, Painlev\'e 
and Chazy equations.
It has been shown \cite{CaGR09} that the use of geometrical techniques to deal with the elements of the Riccati
equation is very efficient to unveil some previously hidden aspects of such equations. The novelty of this article is to study (higher) Riccati equations and various other connected integrable ODEs using coadjoint orbit method  of Virasoro algebra.  In particular, we have given its geometric description using the stabilizer
set of the Virasoro orbit or projective vector field equation. 
 It would be interesting to extend our study to coupled Riccati equations,
how they are connected to stabilizer set of the extended Virasoro orbit or superconformal orbit.
The application of these differential geometric methods to deal with multicomponent systems 
and their integrability is a very interesting subject to be studied.
Hence there are several interesting issues connected to this paper ought to be addressed in future, 
we have only hit the tip of the iceberg. 
We hope to answer some of these questions in forthcoming papers.

%--------------------------------------
\section*{\bf Acknowledgements.}
It is great pleasure to thank Basil Grammaticos, Alfred Ramani, Anindya Ghose
Choudhury and Peter Leach for numerous enlightening discussions and help. J
FC and MFR acknowledge support from research projects MTM-2012-33575, 
and  E24/1 (DGA). 
PG thanks the Departamento de F\'{\i}sica Te\'orica de la Universidad de Zaragoza 
for its hospitality and acknowledges support from IHES, Bures sur Yvette, France 
and CIB, EPFL, Lausanne, Switzerland.

%--------------------------------------

\end{document}